\documentclass[final,3p,times,twocolumn,authoryear]{elsarticle}

\usepackage{amssymb}

\newcommand{\ms}{$\rm m\,s^{-1}$}
\newcommand{\kms}{$\rm km\,s^{-1}$}

\journal{Planetary and Space Science}

\begin{document}
\begin{frontmatter}



\title{Infalling of Nano-dust Because of Air Drag on Uranus}


\author[inst1]{Hua-Shan Shih}

\affiliation[inst1]{organization={Department of Space Science \& Engineering, National Central University},
            addressline={No. 300, Zhongda Rd., Zhongli Dist.}, 
            city={Taoyuan City},
            postcode={320317}, 
            country={Taiwan}}

\author[inst1,inst2]{Wing-Huen Ip}

\affiliation[inst2]{organization={Graduate Institute of Astronomy, National Central University},
            addressline={No. 300, Zhongda Rd., Zhongli Dist.}, 
            city={Taoyuan City},
            postcode={320317}, 
            country={Taiwan}}

\begin{abstract}
Uranus and Saturn share similarities in terms of their atmospheric composition, which is primarily made up of hydrogen and helium, as well as their ring systems. Uranus has 13 known rings, which are divided into narrow main rings, dusty rings, and outer rings. Unlike Saturn's broad ring system, Uranus' inner narrow main rings are relatively narrow, and likely consist of dark, radiation-processed organics that range from centimeters to meters in size. We assume that Uranus may have a mechanism similar to Saturn where tiny particles fall on-to the planet due to its gravity and the dragging force of the upper atmosphere. The uncharged nano-dust particles in Uranus' inner narrow rings will collide with neutral gas molecules in the exosphere and fall onto the planet. This work derives a Monte Carlo simulation of the orbital behavior of nano-dust particles in the inner narrow rings of Uranus. The model shows that the braking of the dust grain motion takes place at altitudes between 6000 km and 8000 km, and the dust particles are gradually captured into corotation with the planetary atmosphere below 4000 km altitude. The larger the dust particles are, the lower the altitude at which they will be assimilated into co-rotation. The lifetime of 1-nm dust particles to 1000 km-altitudes is estimated to be about 32.5 ± 18.8 hours, and that of 30 nm is about 2770.0 ± 213.9 hours.
\end{abstract}





\end{frontmatter}


\section{Introduction} \label{sec:Introduction}

The Uranian ring system is composed of nine narrow rings discovered by stellar occultation measurements in 1977 with the designation of 6, 5, 4, and from inside to outside \citep{Elliot1977, Millis1977, Bhattacharyya1977}. The close Uranus flyby of the Voyager 2 spacecraft added two more rings to the system \citep{Millis1977}, to be followed by the detection of another two rings by the Hubble Space Telescope \citep{Smith1986, Showalter2006}. Overall, the rings are very dark \citep{Smith1986} with some being depleted in dust while others like the lambda ring are quite dusty. From a reanalysis of Voyager 2 images, \citet{Hedman2021} detected the presence of more narrow rings. The size range of the ring particles in the optically thick rings of narrow structures is between cm and m according to \citet{nicholson_pater_french_showalter_2018}.

The origin of Uranian rings has been explained in terms of meteoroid impacts on small satellites and/or mutual collision of ring particles in a system of ringlet belts \citep{Colwell1990}. The narrowness of the rings that can sometimes be as narrow as a few tens of kilometers might be the result of the shepherding effect of small satellites, as presumably in the case of the epsilon ring located between the two 40-km-sized moons, Cordelia and Ophelia \citep{Elliot1984, French1991}. The existence of other moonlets near the alpha and beta rings have been investigated by \citet{Chancia2016}. More recently, \citet{Hearn2022} examined the potential effects of the f-mode resonances generated by the interior oscillations of Uranus on the architecture of the inner rings such as 6, 5, 4, $\alpha$ and $\beta$ without a firm conclusion. 

It was noted by \citet{Broadfoot1986} from the UVS observations during the Voyager 2 flyby of Uranus that the relative lack of microsized dust in the Uranian system might be due to the air drag effect of the extended neutral hydrogen exosphere with a characteristic orbital decay time of less than 1000 years \citep[see also,][]{Colwell1990}. In addition, the Poynting-Robertson effect and the Lorenz force acting on charged dust grains by the planetary magnetosphere can also contribute to the loss of the small dust from the ring system. To some extent, the latter process is reminiscent of the gravitoelectrodynamic injection of charged tiny nanograins from Saturnian rings into Saturn’s upper atmosphere \citep{Northrop1982,Ip1983,Liu2014,Ip2016} or the so-called ring rain mechanism \citep{Donoghue2013} confirmed by the dust experiment on the Cassini mission during its Grand Finale \citep{Hsu2018}. 
What is interesting in the Saturnian ring is that the off-equator ring mass loss rate associated with the charged nanodust amounts to approximately 1800 kg\, s$^{-1}$ to 6800 kg\, s$^{-1}$ \citep{Hsu2018} while the corresponding equatorial mass loss rate because of uncharged nano-dust as a result of atmospheric drag is unexpectedly large (about $10^{4}$ kg\,s$^{-1}$) according to \citet{Mitchell2018}, \citet{Waite2018} and \citet{Perry2018}. It also is suggested that the entry of such a large number of particles into Saturn's atmosphere may have been transient or anomalous \citep{MOSES2023115328}. If the dust influx observed by the Cassini spacecraft is a steady effect, such rapid ring rain equatorial ward would imply a young age of the Saturnian ring system \citep{Crida2019}.

The new discovery of the nanodust transport mechanism in the equatorial plane of Saturn suggests that a similar effect might also take place in the rings of Uranus (and Neptune). The detection of H$_2$O, CO and CO$_2$ in the upper atmosphere of Uranus has been interpreted in terms of the injection of exogenic oxygen-bearing material via meteoroid bombardment, comet impact, and ring dust \citep{Feuchtgruber1997, Cavalie2014, Moses2017, Lara2019}. The mass injection rate, if steady, is on the order of 1--2 kg\,s$^{-1}$ according to the model calculations of \citet{Moses2017} and \citet{Lara2019}. This means that the ring rain process could play an important role at Uranus, since the rings must be subject to constant micro-meteoroid bombardment as well. 

In this work, we apply the Saturnian ring scenario to the case of Uranian rings. We hypothesize a sparse plasma environment within the Uranus ring, where dust particles in nano-meter size , typically exhibit a charge-to-mass ratio ($q/m$) between $10^{-4}$ to $10^{-9}$ e/amu. At an altitude of 10000 km with the Uranus magnetic field around 0.06 to 0.2 Gauss in the equatorial region, the ratio of the Lorentz force to gravity ($F_{\rm Lorentz}/F_{\rm gravity}$) is estimated to be between $10^{3}$ and $10^{-2}$. This suggests a significant influence of the Lorentz force, particularly on tiny dust grains. However, considering the methodologies proposed by \citet{Horanyi1996} and \citet{SZEGO201448}, the estimated charging timescale of the photoemission effect, which represents the duration for a dust grain to charge from 0 to $+$1e, is inefficient for a tiny dust grain to charge, so the influence of the Lorentz force is considered negligible in this study. The organization of the paper is as follows. In Section 2, the basic elements of the dust-air drag orbital calculation is described. The numerical results are analyzed in Section 3. A summary and discussion are given in Section 4.

\section{Method} \label{sec:Method}

\begin{figure}[tbh!]
\centering
\includegraphics[width=1.0\columnwidth]{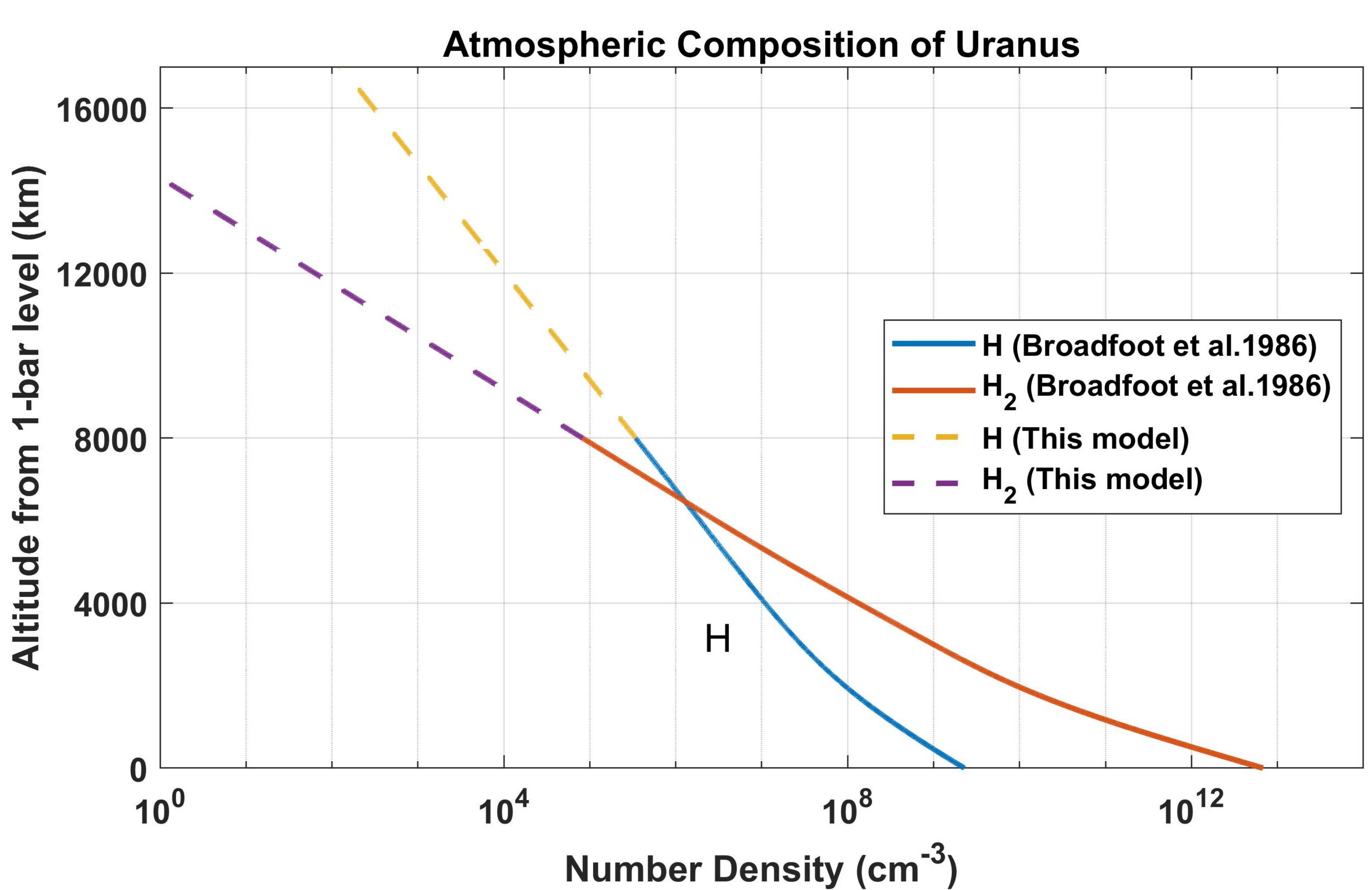}
\caption{The modeling number density profile of hydrogen compounds. The solid lines are models of Voyager 2 UV from \citet{Broadfoot1986}, where atomic hydrogen is an extrapolation of the data measured near 2600 km altitude based on the temperature of 750 from H$_2$. The dashed lines are the estimated number density of the gas with the same slope of the measurements.\label{fig:UranusAtmosphere}
}
\end{figure}

Stellar occultation measurements from the Earth revealed that the light transmission of Uranus’ atmosphere with a minor amount of hydrocarbon molecules was dominated by photo-absorption and Rayleigh scattering of H$_2$ and H atoms \citep{Hudson1971, Mount1977, Mount1978, Smith1990}. Voyager UV occultation measurements of the upper atmosphere of Uranus provided first-hand information on the exospheric temperature ($\sim$800 $\pm$100 K) of the H$_2$ and H gas \citep{Broadfoot1986, Herbert1987, Stevens1993}. The nominal model of the H$_2$ and H number density profiles above the 1-bar pressure level together with the extrapolated curves to be used in the present study is shown in Figure~\ref{fig:UranusAtmosphere}. It is assumed that the exospheric gas particles are in corotation with the central planet with a corotation speed in the azimuthal direction given to be $v_{c}(\rho$) where $\rho$ is the distance perpendicular to the spin axis. If the particle motion is confined to the ring plane, we have $\rho$ = r.

Between two successive collisions, the motion of a dust particle is determined by the gravitational force of the central planet. The average distance traveled between two collisions, the mean-free-path, can be estimated by knowing the corresponding mean-free-paths of collisions with the H$_2$ molecules and H atoms in the exosphere, namely, $\lambda_{\rm H_{2}}$ and $\lambda_{\rm H}$, respectively. If $n_{\rm H_{2}}$ is the number density of H$_{2}$ and $n_{\rm H}$ that of the H atom, and $\sigma$ is the geometrical cross section of the dust particles, $\lambda_{\rm H_{2}}$ = $\frac{v_{\rm th}}{v_{\rm r}}$ $\sigma$ $n_{\rm H_{2}}$ , and $\lambda_{\rm H}$ =$\frac{v_{\rm th}}{v_{\rm r}}$ $\sigma$ $n_{\rm H}$, respectively, where $v_{\rm th}$ is the thermal velocity of the chosen gas particle, and $v_{\rm r}$ =  $v_{\rm c}-v_{\rm d}$  , which is the difference between the local co-rotation velocity ($v_{\rm c}$) and the dust grain velocity ($v_{\rm d}$). The combined collisional mean-free-path is therefore given by,

\begin{displaymath}
\frac{1}{\lambda} =\frac{1}{\lambda_{\rm H}}+\frac{1}{\lambda_{\rm H_2}}
\end{displaymath}

The distance ($S$) travelled by the dust grain before the next collision can then be estimated by generation of a random number ($P_{1}$) between 0 and 1, and finding the value of s by using the following equation,

\begin{displaymath}
P_{1}=1-exp(-\frac{S}{\lambda})
\end{displaymath}

To decide on which gas species the dust grain will hit can be done by computing the ratios of $f_{1}$ = $\frac{\lambda}{\lambda_{\rm H_{2}}}$  and $f_{2}$ = $\frac{\lambda}{\lambda_{\rm H}}$ . We can generate another random number ($P_{2}$) and check whether it is smaller than $f_{1}$. If yes, the dust grain will collide with an H$_2$ molecule. Otherwise, it will collide with an H atom.

The Uranian ring system is quite extended with its outer-most ($\epsilon$) ring at about 51,000 km and inner-most (U2R) ring at about 40000 km. It is likely that dust particles generated by micrometeoroid bombardment or mutual inter-particle collisions will gradually move inward because of ballistic transport \citep[e.g.,][]{Ip1983} or exospheric gas drag effect. This dusty material will finally land at the inner boundary of the ring system defined by the 6 Ring at 41500 km and be injected into the exosphere subsequently. After being released on the ring plane with an initial velocity equal to the Keplerian velocity, the motion of the dust grain will be subject to the planetary gravitational field and the collisional momentum transfer with the exospheric gas.  In our Monte Carlo model of the nano-dust motion, the thermal motion of the exospheric H$_2$ and H is taken into consideration by including the 3D Maxwellian distribution of the thermal velocity into the momentum transfer calculation. As a result, small dust grains of the size of 1--3 nm will experience some amount of random scattering as they spiral downward.

The post-collision velocities of the dust grain (${\bf v_{d}}^{'}$) and of the gas molecule (${\bf v_{g}}^{'}$) can be written as:

\begin{displaymath}
{\bf v_{d}}^{'} = \frac{m_{g}(v_{g}-v_{d})}{m_{d}+m_{g}}+ \frac{m_{d}v_{d}+m_{g}v_{g}}{m_{d}+m_{g}}
\end{displaymath}

\begin{displaymath}
{\bf v_{g}}^{'} = \frac{m_{d}(v_{d}-v_{g})}{m_{d}+m_{g}}+ \frac{m_{d}v_{d}+m_{g}v_{g}}{m_{d}+m_{g}}
\end{displaymath}

where $m_{d}$ is the dust mass, $m_{g}$ can be either the mass of (1) H$_{2}$ or of (2) H, $v_{\rm g}$ is the corresponding gas velocity, and $v_{\rm d}$ is the dust velocity before the collision.


\section{Results} \label{sec:Results}

\begin{figure*}[htb!]
\centering
\includegraphics[width=0.7\textwidth]{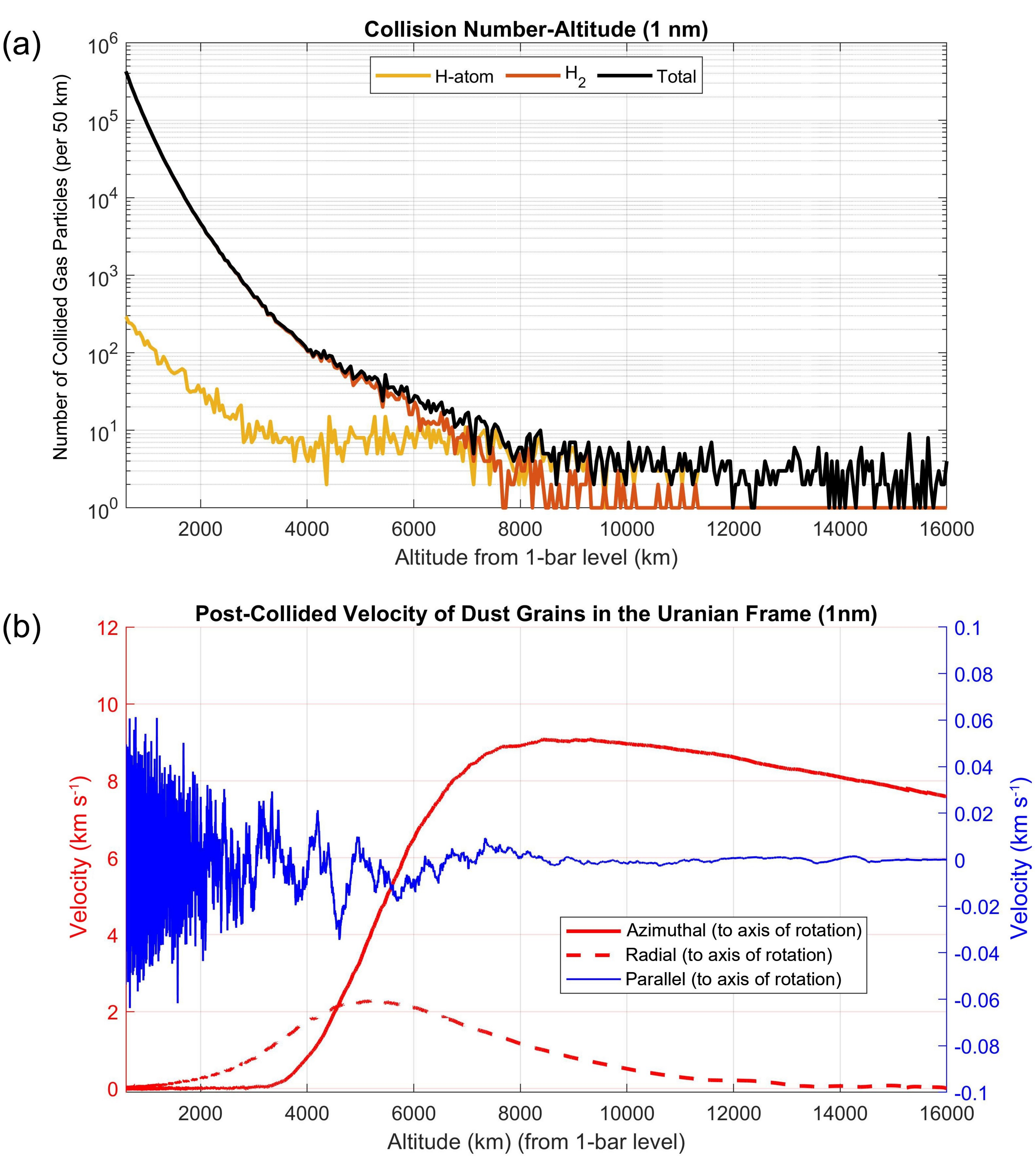} 
\caption{(a) Number distribution the collided hydrogen by the 1 nm-sized dust particles from the altitude of 16,000 km to Uranian lower atmosphere (b) The resulting velocities of the dust grains with the radius of $1 nm$ (mass of about $3800 u$) colliding with atmospheric atoms. \label{fig:UranusCollidN}}
\end{figure*}

\begin{figure*}[htb!]
\centering
\includegraphics[width=0.85\textwidth]{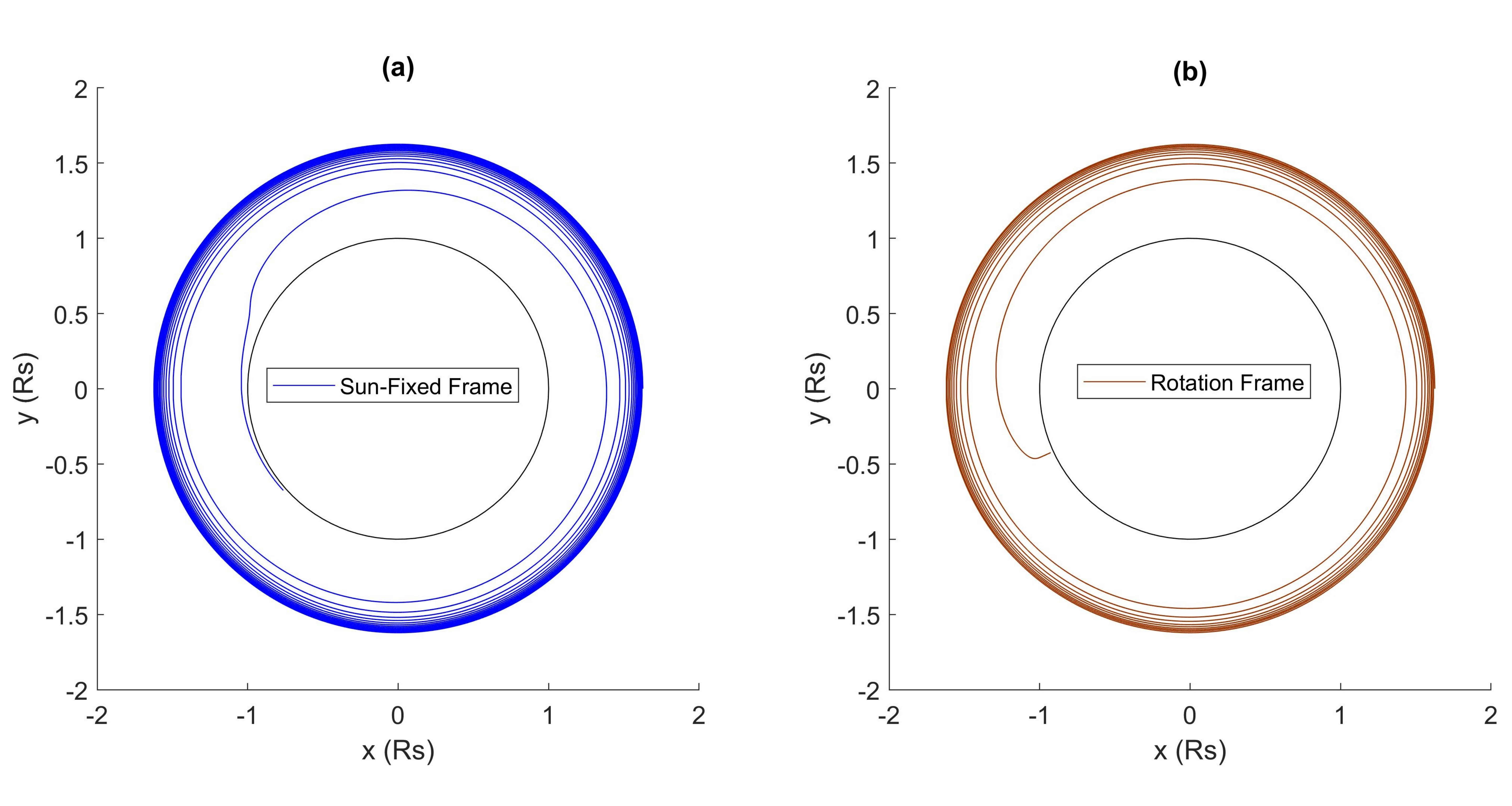} 
\caption{The modelling trajectory profile of a dust particle with the radius of 1 nm colliding with the exospheric gas particles and falling into Uranus center from 6 ring in two frames of reference. \label{fig:Uranus_tra_ab}}
\end{figure*}

Figure~\ref{fig:UranusCollidN}. (a) shows the profiles of H and H$_2$ collision frequency distributions, respectively, of a 1-nm grain after being released from a large ring particle. 
Despite the number density of hydrogen (H) being higher than that of hydrogen molecules (H$_2$) at altitudes above 8000 km, the collision frequency for both with the single dust particle is low due to the sparse nature of the atmosphere at such high altitudes, resulting in less than 10 collisions per 50 km. However, as the dust particle descends to altitudes around 4000 km, the collision frequency increases significantly, with more than 100 collisions per 50 km occurring because of the increased density of the atmosphere.
A significant part of the braking of the dust grain motion takes place between these two altitudes.  It is also interesting to note that because of collisional scattering by the exospheric H$_2$ molecules with thermal motion, the 1-nm nano-dust grain could have a small random speed ($<$ 75 \ms{}) at the end of its inward spiral. As in analyses of the Cassini dust measurements during its Grand Finale \citep{Mitchell2018}, the orbital evolution of the nano-dust particles can be viewed either in the Sun-fixed frame or the rotating frame of the planet. Figure~
\ref{fig:Uranus_tra_ab} show the trajectory of an uncharged dust grain with the radius of 1 nm in (a) the Sun-fixed frame  and (b) Uranus’ co-rotating frame. The dust grain slowly lowers its altitude due to random collisions with the exospheric gas particles. 

Figure~\ref{fig:UranusCollidN}. (b) shows how, at first, the total speed of a 1-nm dust grain increases because of gravitational acceleration. But the trend reverses and the inward orbital spiraling speeds up as collisions become more frequent until finally being captured into full co-rotation with the planetary atmosphere at about 4000 km altitude. While the initial speed of 11.81 \kms{} from circular Keplerian motion is higher than the co-rotation speed of 4.21 \kms{}, the azimuthal velocity component will gradually increase before being reduced as a consequence of the gas drag effect. As the dust falls below the exobase at about 6500 km--7000 km altitude, the number density of the atmospheric neutral gas is relatively high. The probability of particle collision increases greatly, causing the azimuthal velocity component to be strongly decelerated. The relative velocity of the dust and Uranus’ atmosphere will eventually become close to zero at about 3000 km altitude. The radial velocity component of the dust particle also increases of a maximum value of 2.3 \kms{} before being slowed down to nearly zero subsequently (Figure~\ref{fig:UranusCollidN}. (b), dashed red line). 

\begin{figure*}[tbh!]
\centering
\includegraphics[width=0.6\textwidth]{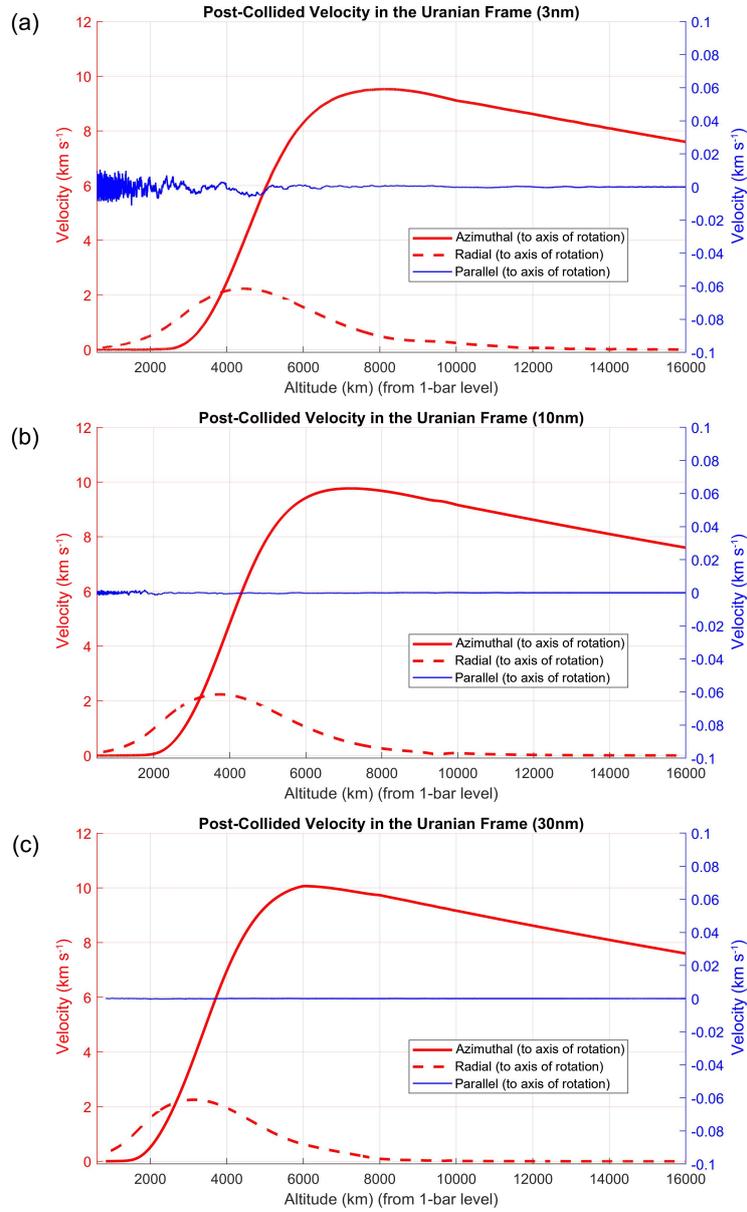}
\caption{The resulting velocity changes of the dust grains with the radius of (a) $3 nm$, (b) $10 nm$, and (c) $30 nm$
\label{fig:V_abc}}
\end{figure*}

Figure~\ref{fig:V_abc} (a)--(c) compare the dynamical behaviors of dust grains of three other radii, i.e., 3 nm, 10 nm and 30 nm. It can be seen that they generally follow the same pattern. The only difference is that the larger the dust grains, the lower will be the altitude for them to be totally assimilated into co-rotation of the planetary atmosphere. 

\begin{figure*}[tbh!]
\centering
\includegraphics[width=0.7\textwidth]{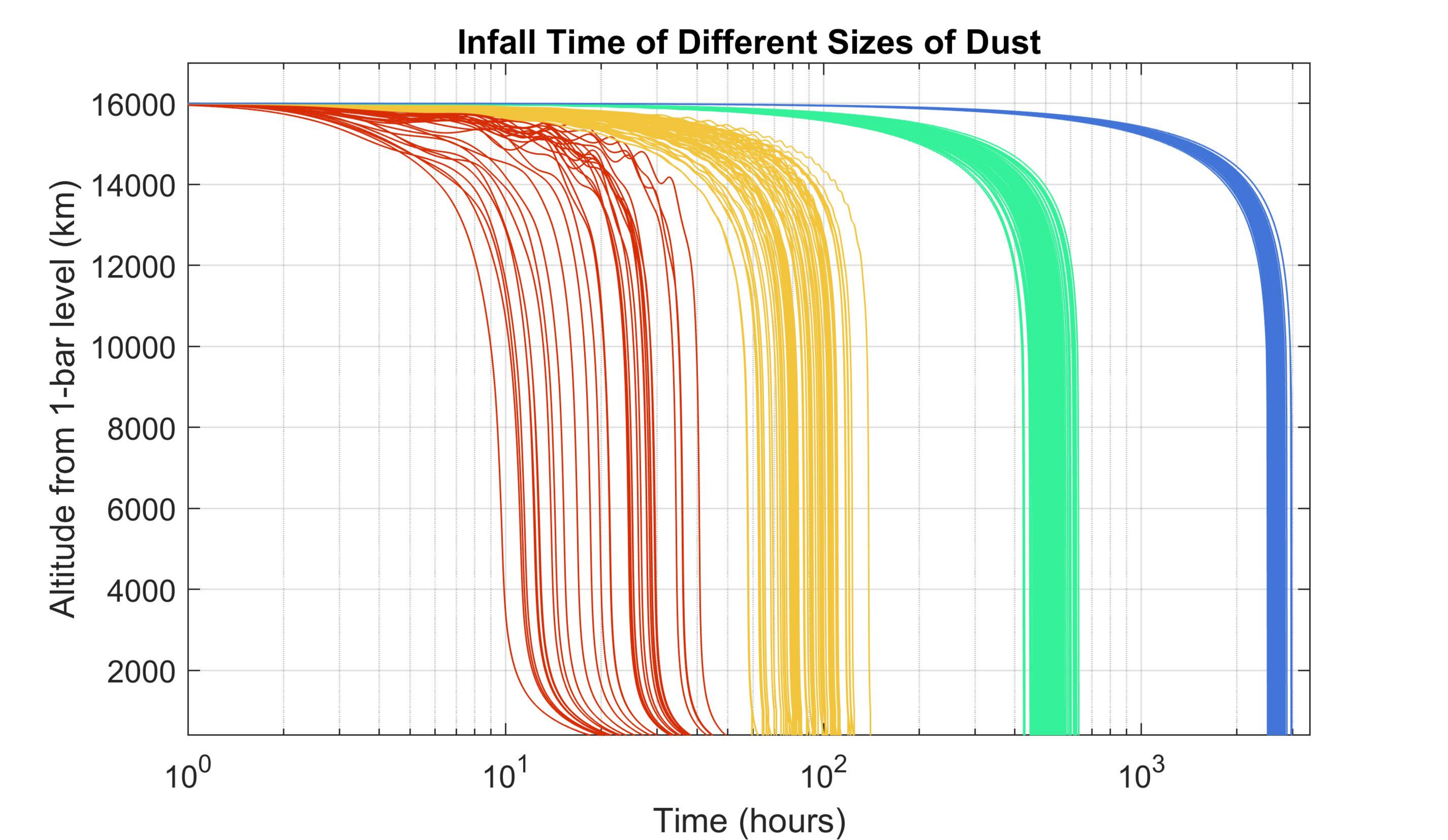}
\caption{Comparison of the drop-off time of (red) $1 nm$, (yellow) $3 nm$, (green) $10 nm$, and (blue) $30 nm$  of dust falling into Uranus.
\label{fig:Time}}
\end{figure*}

Finally, the time dependence of the dust grains is summarized in Figure~\ref{fig:Time}. Examples of the four cases with different grain sizes are shown. In our Monte Carlo model calculations, hundreds simulation runs have been performed. The results show that the orbital histories of small grains with size $\lesssim$ 3 nm are highly dispersed because their motion can be significantly influenced by the thermal motion of the gas particles. On the other hand, larger dust grains experienced lesser changes in their motion. Further analysis of the simulation results revealed that the infalling time of 1-nm dust particles was dry short, with a duration on the order of 32.5 ± 18.8 hours. During this time, the dust particles mostly rotate at an altitude of 12,000 km for 10 to 35 hours, before spiraling below 2000 km for 7--8 hours. In comparison, 30-nm dust particles will take 2770.0 ± 213.9 hours to reach a lower altitude. If dust grains with a radius of 30 nm reach altitudes below 10,000 km, they would fall to an altitude of 1,000 km above Uranus in only 20 hours. The test of dust with other sizes showed that the lifetime of 3 nm was 97.0 ± 213.9 hours, and that of 10 nm was 525.2 ± 94.7 hours.

\section{Summary and Discussion} \label{sec:Discussion}

Studying the behavior of nano-dust particles in the exosphere of Uranus can provide valuable insights into the inward transport of dusty material from the Uranian ring system. This study suggests that as the uncharged dust grains with different radii, ranging from 1 nm to 30 nm spirals inward, the collision frequency of dust particles with hydrogen atoms and hydrogen molecules increases. The trajectory of a 1-nm dust grain slowly lowers its altitude because of random collisions with the exospheric gas particles. The gravitational acceleration initially increases the dust grain's total velocity, but the trend reverses as collisions become more frequent. The inward orbital spiraling also speeds up until the grain is captured into co-rotation with the Uranian atmosphere below the altitude of 4000 km. The collisional scattering by the exospheric molecules with thermal motion gives the nano-dust grain a small random speed 0.075 \kms{} at a lower altitude of Uranus. The dynamical behaviors of dust grains with radii of 3 nm, 10 nm, and 30 nm were also analyzed. The results revealed that the trajectories of small dust grains with a radius less $\lesssim$ 3 nm are characterized by a high degree of dispersion due to the significant influence of thermal motion of exospheric gas particles on their dynamics, while larger dust grains experience lesser changes in their motion. The infalling time of 1-nm dust particles is very short, with a duration of about 32.5 ± 18.8 hours, and larger dust grains take longer to reach lower altitudes. The study also highlights the effect of gas drag on the azimuthal velocity component of the dust particles, which slows them down as they fall below the exobase. 

Our toy-model calculation described here is meant to show the possible dynamical evolution of the nano-size dust grains ejected from the inner ring region. It shows that the extended exosphere of Uranus can be very effective in drawing in the oxygen-bearing material thus providing a source of the H$_2$O, CO and CO$_2$ detected in the upper atmosphere of Uranus \citep{Moses2017, Lara2019, Feuchtgruber1999}. As in the case of the Saturnian rings, the corresponding dust infalling rates, hopefully, to be measured by a Uranus Orbiter \citep{Cohen2022}, would provide constraints on the lifetime and hence origin of the ring system. An interesting issue which is outside the scope of the present work is about the dynamics of charged dust particles. Because of the highly offset and tilted magnetic field configuration, sub-micron size particles emitted from the Uranian ring system could be transported to a wide region of the planetary magnetosphere. Altogether, the Uranian ring system could have interesting connections to the composition of the upper atmosphere and magnetospheric processes, in spite of its tenuous nature. Overall, the study provides a framework for future research in this area. It demonstrates the interaction between exospheric gas particles and dust particles, and highlights the need for further investigation into the processes driving the inward transport of dusty material from the Uranian ring system. A similar dust infall process is also likely to take place in the vicinity of the Neptunian ring system.

\section*{Acknowledgements}

We are thankful for the comments and suggestions from the anonymous referees to help improve the quality of this paper.
This work was supported by the National Science and Technology Council (NSTC, Taiwan), grant NO. 111-2112-M-008-014-. 



\bibliographystyle{elsarticle-harv} 
\bibliography{master}





\end{document}